\documentstyle[prl,aps,multicol,epsf]{revtex}
\begin{document}
\draft
\title{Ground-state phase diagram of the one-dimensional dimerized 
$t$$-$$J$ model \\at quarter filling}
\author{S. Nishimoto$^1$ and Y. Ohta$^{1,2}$}
\address{$^1$Graduate School of Science and Technology, 
Chiba University, Inage-ku, Chiba 263-8522, Japan\\
$^2$Department of Physics, Chiba University, Inage-ku, 
Chiba 263-8522, Japan} 
\date{May 10,1998}
\maketitle
\begin{abstract}
The ground state of the one-dimensional dimerized $t$$-$$J$ 
model at quarter filling is studied by a Lanczos exact-diagonalization 
technique on small clusters.  We calculate the charge gap, spin gap, 
binding energy, Tomonaga-Luttinger-liquid parameter, Drude weight, 
anomalous flux quantization, etc.  We thereby show that the two 
types of dimerization, i.e., a dimerization of hopping integral 
and a dimerization of exchange interaction play a mutually 
competing role in controlling the ground state of the system 
and this leads to the emergence of various phases including 
the Mott insulating, Tomonaga-Luttinger-liquid, and 
Luther-Emery-liquid phases.  The ground-state phase diagram of 
the model is given on the parameter space of the dimerizations.  
\end{abstract}
\begin{multicols}{2}

\section{INTRODUCTION}

There are a number of low-dimensional correlated electron systems 
described by the Hubbard and $t$$-$$J$ models at quarter-filling.  
One of the examples is the Bechgaard salts (TMTSF)$_2$X and 
(TMTTF)$_2$X with X=PF$_6$, ClO$_4$, Br, etc., where there are 
three electrons in the two highest-occupied molecular-orbitals of a 
dimerized molecule, e.g., (TMTTF)$_2$, and the system is at 
$\frac{3}{4}$ filling in terms of electrons, which corresponds 
to the quarter filling in terms of holes\cite{[1]}.  In this system, 
dimerization of the molecules is known to play an essential role: 
depending primarily on the strength of dimerization, there appears 
a variety of electronic phases, such as antiferromagnetic insulating, 
paramagnetic metallic, and superconducting phases\cite{[1]}.  Some 
theoretical calculations of the dimerized Hubbard models have been 
done to clarify the nature of this system\cite{[2],[3]}.  

Another example is the transition-metal oxide NaV$_2$O$_5$ which 
is reported to be a quarter-filled ladder system, exhibiting a 
spin-Peierls--like phase transition accompanied by a charge 
ordering\cite{[4]}.  This system may also be regarded as a dimerized 
system at quarter filling if we may assume that two V ions on 
each rung of the ladder form a dimer molecule\cite{[5]}.  
Effects of the lattice dimerization on the correlated electron 
systems in two dimension (2D) have also been studied in connection 
with cuprate superconductivity, where the spin-gap phenomena above 
$T_c$ have attracted much attention.  The simplest possible model 
that exhibits a spin gap which can survive against hole doping may 
be a dimerized $t$$-$$J$ model with an alternating exchange 
interaction, where an enhancement of the singlet 
superconducting correlation has been suggested\cite{[6]}.  

Motivated by such developments in the field, we study in this paper 
the one-dimensional (1D) dimerized $t$$-$$J$ model at quarter 
filling, of which not very much is known so far.  
The model we study is defined by the Hamiltonian
\begin{eqnarray}
&H&=-t_1\sum_{\langle ij \rangle\sigma}
(\hat{c}^\dagger_{i\sigma}\hat{c}_{j\sigma}+{\rm H.c.})
-t_2\sum_{\langle kl \rangle\sigma}
(\hat{c}^\dagger_{k\sigma} \hat{c}_{l\sigma}+{\rm H.c.})
\nonumber \\
&+&J_1\sum_{\langle ij \rangle}({\bf S}_i\cdot{\bf S}_j
-\frac{n_in_j}{4})
+J_2\sum_{\langle kl \rangle}({\bf S}_k\cdot{\bf S}_l
-\frac{n_kn_l}{4})
\end{eqnarray}
where $\hat{c}^\dagger_{i\sigma}$$=$$c^\dagger_{i\sigma}(1$$-$$n_{i-\sigma})$ 
is the constrained electron-creation operator at site $i$ and spin $\sigma$ 
$(=\uparrow,\downarrow)$, ${\bf S}_i$ is the spin-$\frac{1}{2}$ operator, 
and $n_i$ is the electron-number operator; hereafter we refer to the 
fermionic particle as `electron', which corresponds to, e.g., the hole in 
the organic compounds.  
We use the 1D lattice shown in Fig.~1; $\langle ij\rangle$ stands for 
nearest-neighbor bonds with parameters $t_1$ and $J_1$ and $\langle kl\rangle$ 
for those with parameters $t_2$ $(\ge$$t_1)$ and $J_2$ $(\ge$$J_1)$.  
The model tends to the usual $t$$-$$J$ model when there is no dimerization 
($t_1$$=$$t_2$ and $J_1$$=$$J_2$), of which much study has been 
made\cite{[7],[8],[9],[10]}, whereas in the limit of strong dimerization, 
the model represents an assembly of isolated dimers.  
We retain the relations between parameters $t$ and $J$ obtained from 
perturbation, i.e., $J_1$$=$$4t^2_1/U$ and $J_2$$=$$4t^2_2/U$, 
in order to reduce the number of parameters, where $U$ is the 
corresponding on-site Hubbard interaction.  
We thereby keep a relation $J_1/J_2$$=$$(t_1/t_2)^2$.  
We thus have three independent parameters, and if we take $t_2$ 
as a unit of energy, then we have two parameters, for which we will take 
parameters representing $t$-dimerization and $J$-dimerization (of which 
a specific definition is given below).  

We employ a Lanczos exact-diagonalization technique\cite{[11]}, which is 
used to obtain energies of the ground state and a few low-lying excited 
states.  We denote the number of lattice sites by $N_s$ and the numbers 
of up- and down-spin electrons by $N_\uparrow$ and $N_\downarrow$, respectively.  
The electron density is then given by $n=(N_\uparrow+N_\downarrow)/N_s$, so 
that $n$$=$$1$ represents the half-filled system.  Here we restrict ourselves 
to the case of $n$$=$$1/2$, i.e., quarter filling.  We use the finite-size 
systems of the size 4, 6, and 8 unit cells (or 8, 12, and 16 sites, 
respectively).  In order to achieve a systematic convergence to the 
thermodynamic limit, we choose periodic boundary condition for 
$N_\uparrow+N_\downarrow=4m+2$ and antiperiodic boundary condition for 
$N_\uparrow+N_\downarrow=4m$ where $m$ is an integer.  

We will examine the ground-state properties by calculating the charge gap, 
spin gap, Luttinger liquid parameters, anomalous flux quantization, etc., 
as a function of the dimerization strength.  
We will thereby show that the competition between the two types of 
dimerization leads to various ground-state phases such as a Mott 
insulating phase, Tomonaga-Luttinger-liquid phase, spin-gap phase, 
etc.  The ground-state phase diagram of the model is thereby given.  

This paper is organized as follows; 
In Sec.~II, we present the calculated charge and spin gaps and clarify 
the mechanism of insulator-metal (or superconductor) transition.  
In Sec.~III, we calculate the Tomonaga-Luttinger-liquid properties of the 
model and also discuss a possibility of singlet superconductivity.  
In Sec.~IV we present a phase diagram of the system by summarizing 
the results given in Sec.~III and IV.  Conclusions are given in Sec.~V.  

\section{INSULATOR-METAL TRANSITION}

In this section, we calculate the charge gap, spin gap, and binding 
energy of the model, and discuss the mechanism of insulator-metal 
(or superconductor) transition of the system.  
First let us introduce two parameters for dimerization; i.e., 
the strength of dimerization in the hopping integral defined as 
\begin{equation}
\tilde{t}_d=\frac{t_2-t_1}{t_1}
\end{equation}
which we call $t$-dimerization, and the strength of dimerization in 
the exchange interaction defined as 
\begin{equation}
\tilde{J}_d=\frac{J_2-J_1}{t_1}
\end{equation}
which we call $J$-dimerization.  We also take $J_2/t_2$ as a measure 
of the strength of $J$-dimerization because if we keep $\tilde{t}_d$ 
constant $(>$$0)$ then $J_2/t_2\propto \tilde{J}_d$.  
These are the key parameters which control the electronic state of the 
system; the $t$-dimerization has the effect leading to the repulsive 
interaction among electrons that acts when different spins come in a 
single dimer, and the $J$-dimerization has the effect promoting the 
spin-singlet formation between spins coming in a single dimer.  
These effects manifest themselves in the electronic state of an isolated 
$t$$-$$J$ dimer with an electron.  The single-particle gap of the dimer 
is given as $U_{\rm dimer}$$=$$2t$$-$$J$, where $t$ and $J$ are the hopping 
and exchange parameters of the single bond, and thus $U_{\rm dimer}$ may 
be regarded as the effective Hubbard-like interaction of the dimer.  The 
value of $U_{\rm dimer}$ can be either positive or negative depending 
on whether the value $J/t$ is smaller or larger than 2.  The competing 
effects of the two types of dimerization may thus be explained in the 
limit of strong dimerization.  Now let us examine whether the competition 
between the effects of two types of dimerization in the 1D system can 
lead to the insulator-metal (or superconductor) transition.  For this 
purpose we calculate the charge gap, spin gap, and binding energy.  

\subsection{Charge gap}

The charge gap may be defined by 
\begin{eqnarray}
\Delta_c=\frac{1}{2}\Big[\big[E_0(N_\uparrow+1,N_\downarrow)
&-&E_0(N_\uparrow,N_\downarrow)\big]   \nonumber \\
-\big[E_0(N_\uparrow,N_\downarrow)
&-&E_0(N_\uparrow-1,N_\downarrow)\big]\Big] 
\end{eqnarray}
where $E_0(N_\uparrow,N_\downarrow)$ is the ground-state energy 
of the system with $N_\uparrow$ up-spin and $N_\downarrow$ down-spin 
electrons.  This is evaluated for a number of clusters; throughout 
the paper we use a linear extrapolation of the data obtained from 
finite-size clusters with respect to $1/N_s$ to estimate the value 
at the infinite system size because to use more sophisticated scaling 
formula will give no qualitative difference to the obtained results 
although unphysical small negative values sometimes appear due to 
the errors of this scaling.  The calculated results for $\Delta_c$ 
as a function of the $t$- and $J$-dimerizations (i.e., $\tilde{t}_d$ 
and $J_2/t_2$) are shown in Fig.~2 (a).  

In case where there is no dimerization ($t_1$$=$$t_2$ and 
$J_1$$=$$J_2$), i.e., for the homogeneous $t$$-$$J$ model, a number 
of studies\cite{[7],[8],[9],[10]} have shown that there is no charge gap 
$\Delta_{c}/t_2$$=$$0$ in the entire region of $J_2/t_2$ 
(except in the region of phase separation).  Our result is 
consistent with this.  
In case where there is dimerization, the charge gap opens as seen 
in Fig.~2 (a).  
At $J_2/t_2$$\rightarrow$$0$ the system is equivalent to the 
1D dimerized Hubbard model at $U$$\rightarrow$$\infty$, so that 
the dimerization gap of the size $\Delta_D=2(t_2$$-$$t_1)$ opens 
at the Brillouin zone boundary, which is the charge gap.  
In other words, the separation between the lowest unoccupied 
(antibonding) band and the highest completely occupied (bonding) 
band is given by $\Delta_D$.  In real-space picture, there is one 
electron per dimer, i.e., the number of electrons is equal to 
the number of dimers, and by regarding the dimer as a site one 
may have an effective half-filled band with the effective Coulomb 
repulsion of $U_{\rm eff}=\Delta_D$.  
We thus have a Mott insulator due to $t$-dimerization.  

With increasing $J$-dimerization $J_2/t_2$, we find that $\Delta_{c}/t_2$ 
decreases and becomes zero at a value $J_2/t_2$$=$$(J_2/t_2)_c^{\rm charge}$ 
where the gap closes.  $(J_2/t_2)_c^{\rm charge}$ represents the 
critical strength of $J$-dimerization at which a transition from 
insulating phase to metallic (or superconducting) phase occurs.  
This may be explained as follows; 
In the insulating phase ($\Delta_c/t_2$$>$$0$), the effective repulsion 
$U_{\rm eff}$ is given by the difference between loss of the 
kinetic energy and gain in the exchange interaction when one brings 
two electrons into a dimer, i.e., $U_{\rm eff}$$=$$2(t_2$$-$$t_1)$$-$$J_2$.  
Thus, with increasing $J_2/t_2$, the measure of the gap $U_{\rm eff}$ 
decreases, and at some value of $J_2/t_2$ the charge gap closes.  
In the small $J$-dimerization limit, we may use the perturbation 
theory with respect to $J_2/t_2$ (or $J_1/t_1$)\cite{[3]}.  After a 
straightforward calculation we obtain the result for the 
charge gap:
\begin{equation}
\Delta_{c}=2(t_2-t_1)\Big[1-\frac{\ln2}{\pi}
\Big(\frac{J_1}{t_1}+\frac{J_2}{t_2}\Big)
\ln\frac{4(t_1+t_2)}{t_2-t_1}\Big]. 
\end{equation}
In the limit of $J_2$$\rightarrow$$0$ and $J_1$$\rightarrow$$0$, this 
reduces to $\Delta_c$$=$$2(t_2-t_1)$, which is equal to $\Delta_D$ 
defined above.  We compare the exact-diagonalization data with this 
analytical expression in Fig.~2 (b) where we find a good agreement.  

With increasing $J_2/t_2$ further, we find that at some $J_2/t_2$ 
value the charge gap starts to increase again (see Fig.~2 (a)), at 
which the spin gap opens as we will see below.  This is because 
$\Delta_{c}$ reflects the effect of singlet binding energy in the 
charge-gapless region.  

\subsection{Spin gap and binding energy}

The spin gap may be defined by 
\begin{eqnarray}
\Delta_s=E_0(N_\uparrow+1,N_\downarrow-1)
-E_0(N_\uparrow,N_\downarrow).
\end{eqnarray}
The calculated results for $\Delta_s$ as a function of the $t$- and 
$J$-dimerizations are shown in Fig.~3.  When there is no dimerization, 
the results obtained are consistent with a recent result for the 
$t$$-$$J$ model\cite{[10]}.  
In a small $J_2/t_2$ limit we have $\Delta_s$$\rightarrow$$0$ because 
there is no spin correlations in the system.  
With increasing $J_2/t_2$ with a fixed strength of $t$-dimerization, 
we find that the spin gap opens at some $J_2/t_2$ value, which we define 
as $(J_2/t_2)_c^{\rm spin}$, the critical strength of $J$-dimerization 
at which the spin gap opens.  
We find that the spin gap remains finite in the region between 
$J_2/t_2$$=$$(J_2/t_2)_c^{\rm charge}$ and the larger $J_2/t_2$ value 
at which the phase separation occurs.  
We also find that with increasing $\tilde{t}_d$, the critical 
strength $(J_2/t_2)_c^{\rm spin}$ becomes smaller and at the same 
time $\Delta_s/t_2$ increases.  
This means that the spin gap is enhanced by increasing 
the $t$-dimerization.

We note that the relation 
$(J_2/t_2)_c^{\rm spin}>(J_2/t_2)_c^{\rm charge}$ always holds.  
This suggests that there exist two types of metallic region in the 
model; one is the phase where there are both gapless spin and 
gapless charge modes, and the other is the phase where there is 
a gap only in the spin excitation, which are the so-called 
Tomonaga-Luttinger (TL) region and Luther-Emery (LE) region, 
respectively.  We will discuss this in Sec.~III.  

A simple picture may be given to the case of strong $J$-dimerization 
where the spin-gap formation is ensured.  
When even number of electrons exist, in the lowest order 
perturbation of $t_2/J_2$ two electrons with opposite spins 
always make a pair on the dimerized bond and gain the singlet 
formation energy $3J_2/4$.  The electrons hop only as a pair 
tunneling through the virtual pair breaking in the forth 
order of $t_1/J_1$ or $t_2/J_2$.  
The effective Hamiltonian is of the form 
\begin{eqnarray}
H_{\rm eff}=-\tilde{t}(s_i^\dagger s_{i+2}+{\rm H.c}),
\end{eqnarray}
where 
\begin{eqnarray}
s_i^\dagger=\frac{1}{\sqrt{2}} 
(c_{i\uparrow}^\dagger c_{i+1\downarrow}^\dagger 
-c_{i\downarrow}^\dagger c_{i+1\uparrow}^\dagger)
\end{eqnarray}
is the singlet Bose operator at site $i$ (even number) and 
\begin{eqnarray}
\tilde{t}=\Big(2+\frac{t_1^2}{t_2^2-t_1^2}\Big)\,
\frac{t_1^2\,t_2^2}{J_2^3}
\end{eqnarray}
is the effective hopping parameter of the boson.  
The Hamiltonian Eq.~(7) may be derived in the same way as the 
effective Hamiltonian of the attractive Hubbard model in the 
strong-coupling limit\cite{[6],[12]}.  

We also calculate the binding energy defined as
\begin{eqnarray}
\Delta_{B}=\big[E_0(N_\uparrow&+&1,N_\downarrow-1)
-E_0(N_\uparrow,N_\downarrow)\big]   \nonumber \\
&-&2\big[E_0(N_\uparrow+1,N_\downarrow)
-E_0(N_\uparrow,N_\downarrow)\big], 
\end{eqnarray}
which is negative if two electrons minimize their energy by producing 
a bound state, and indicates a possible superconductivity.  This 
value is meaningless unless the system is metallic or 
$J_2/t_2$$>$$(J_2/t_2)_c^{\rm charge}$.  
In Fig.~4, we show calculated results for $\Delta_B$ as a function of
$J_2/t_2$, where we find $\Delta_B$$\simeq$$0$ for all values of 
$\tilde{t}_d$$>$$0$ until $J_2/t_2$$=$$(J_2/t_2)_c^{\rm spin}$ is 
reached, and at this critical point $\Delta_B$ starts to decrease 
suddenly.  This mean that a bound state of two holes is formed in the 
entire region of the spin-gap phase and the binding energy is enhanced 
as $J_2/t_2$ is increased.  We also note that at constant $J_2/t_2$ 
the value of $\Delta_B$ increases with increasing of $\tilde{t}_d$; 
i.e., the singlet binding energy is enhanced by the $t$-dimerization.  

\section{LUTTINGER LIQUID BEHAVIOR}

It is well known that the 1D interacting fermion systems may be 
related to the Fermi-gas model in the continuum limit, where there 
are two different regimes, the Tomonaga-Luttinger (TL) regime and 
Luther-Emery (LE) regime.  The essential difference between the two 
lies in the spin degrees of freedom; the TL region is characterized 
by the liquid phase with both a gapless spin mode and a gapless 
charge mode, whereas in the LE region the charge degrees of freedom 
is described by the TL liquid but the spin degrees of freedom 
have a gap.  According to the TL liquid theory\cite{[13],[14],[15]}, 
various combinations of the parameters $K_\rho$ and $K_\sigma$ 
describe the critical exponents of correlation functions of the 
system.  In the absence of magnetic field, the dimerized $t$$-$$J$ 
model is isotropic in spin space, so that $K_\sigma$$=$$1$ holds, 
and we are left with the only parameter $K_\rho$.  

In the TL region, the spin and charge correlation functions show 
a power-law dependence as
\begin{eqnarray}
\langle S_i^zS_{i+r}^z\rangle\sim e^{2ik_{\rm F}r}\big/r^{K_\rho+K_\sigma}
\end{eqnarray}
and
\begin{eqnarray}
\langle n_in_{i+r}\rangle\sim e^{2ik_{\rm F}r}\big/r^{K_\rho+K_\sigma}
+e^{4ik_{\rm F}r}\big/r^{4K_\rho}, 
\end{eqnarray}
respectively.  
We see that, for $K_\rho$$<$$1$, 2$k_{\rm F}$-SDW or 2$k_{\rm F}$-CDW 
are enhanced and diverged, whereas for $K_\rho$$>$$1$, pairing 
fluctuations dominate.  
In the LE region, on the other hand, the spin gap opens and the 
2$k_{\rm F}$-SDW correlation decays exponentially.  Because the 
contribution from spin excitations vanishes, the critical exponent 
of 2$k_{\rm F}$-CDW also changes and the asymptotic form is 
given by
\begin{eqnarray}
\langle n_in_{i+r}\rangle\sim e^{2ik_{\rm F}r}\big/r^{K_\rho}.  
\end{eqnarray}
Thus, in the region with $K_\rho$$>$$1$, the singlet spin correlation 
dominates over the 2$k_{\rm F}$-CDW correlation.

The relations between the correlation exponent $K_\rho$ and the 
low-energy behavior of the model given below are useful for the 
evaluation of $K_\rho$;  First, the parameter $K_\rho$ is obtained 
from the charge compressibility $\kappa$ and charge velocity $v_c$ as 
\begin{eqnarray}
{K_\rho}=\frac{\pi}{2}v_c\kappa.
\end{eqnarray}
The charge velocity may be determined by 
\begin{eqnarray}
v_c = \frac{N_s}{2\pi}(E_{1,S=0}-E_0)  
\end{eqnarray}
where $E_{1,S=0}$ is the energy of the lowest charge mode (measured 
from the ground-state energy $E_0$) at a neighboring $k$ point.  
The inverse compressibility is given by 
\begin{eqnarray}
\frac{1}{n^2\kappa}&=&\frac{1}{N_s}\frac{\partial^2E_0}{\partial n^2} \nonumber\\
&=&\frac{2}{N_s}\Bigg[\big[E_0(N_\uparrow+2,N_\downarrow)
-E_0(N_\uparrow,N_\downarrow)\big]                              \nonumber\\
&~&~~~~~~~~~-\big[E_0(N_\uparrow,N_\downarrow)-E_0(N_\uparrow-2,N_\downarrow)
\big]\Bigg]^{-1}.
\end{eqnarray}
The parameter $K_\rho$ is also related to the Drude weight $D$, the weight 
of the zero-frequency peak in the optical conductivity $\sigma_\omega$, and 
may be obtained by considering the curvature of the ground-state energy-level 
as a function of the threaded flux\cite{[16],[17],[18],[19]}: 
\begin{eqnarray}
D=2v_cK_\rho =\frac{N_s}{4\pi}\frac{\partial^2(E_0)}{\partial \Phi^2}.
\end{eqnarray}
Equations (14)--(17) provide us with independent conditions 
on $K_\rho$, $v_c$, and $D$, which can be used to evaluate the TL-liquid 
parameter and to check the consistency in the TL-liquid relations.  

\subsection{Tomonaga-Luttinger-liquid parameter}

The calculated results for the TL-liquid parameter $K_\rho$ are 
shown in Fig.~5 (a) as a function of $J_2/t_2$ for various strength of 
$t$-dimerization where the 8-, 12-, 16-site clusters are used although 
the cluster-size dependence of $K_\rho$ is small.  

A limiting case of $J_2/t_2$$\rightarrow$$0$ may be considered first.  
The ground state can be obtained by using the first-order degenerate 
perturbation theory around $J_1$$=$$J_2$$=$$0$, where the wave function 
is the same as that in the $U/t$$\to$$\infty$ dimerized Hubbard model, 
i.e., a product of the state of spinless fermions describing the 
charge degrees of freedom localized at each dimer and the state 
describing the spin system.  The 2$k_{\rm F}$-SDW correlation is 
dominant here: there is no chance for superconductivity.  
Note that even if the additional three-site terms are present, 
which are obtained through the Schrieffer-Wolff transformation to 
derive the dimerized $t$$-$$J$ model from the dimerized Hubbard 
model, they do not bring any change in the wave function in the 
first-order perturbation.  Consequently, the dimerized $t$$-$$J$ 
model in the $J_2/t_2$$\to$$0$ limit for any $\tilde{t}_d$ values 
gives us the value of $K_\rho$$=$$1/2$.  We however note that, at 
$J_2/t_2$$=$$0$ with a finite $t$-dimerization, the charge gap 
is positive $\Delta_c/t_2$$>$$0$ and the system is a Mott insulator.  

With increasing $J_2/t_2$, at any $\tilde{t}_d$, $K_\rho$ 
increases, and in the intermediate strength of $J_2/t_2$ (and for 
a rather small $\tilde{t}_d$), there appears the region with 
$K_\rho$$>$$1$ where the superconducting correlation is the most 
dominant.  
In the $g$-ology, this means the region of $g_2$$<$$0$ is effectively 
realized due to the attraction interaction caused by the 
$J$-dimerization.

In Fig.~5 (b), we show the contour map for $K_\rho$ on the parameter 
space $(t_1/t_2,J_2/t_2)$ where the contour lines are drawn by using 
a spline interpolation.  We find that, for a fixed $J_2/t_2$, $K_\rho$ 
decreases as $t$-dimerization increases.  But as shown above the spin 
gap are enhanced by the increase of $t$-dimerization, so that the 
phase with both $\Delta_s$$>$$0$ and $K_\rho$$>$$1$ is realized in a 
reasonably wide range of the parameter values.  We thus confirm that 
the singlet superconducting phase indeed exists.  
We also find that the calculated spin and charge correlation 
functions fit very well with Eqs.~(11) and (12) as seen in Fig.~5 (b).  

\subsection{Drude weight and flux quantization}

The calculated results for the Drude weight defined in Eq.~(17) are 
shown in Fig.~6.  We find that, as $t$-dimerization increase, the 
dependence of $D$ on $J_2/t_2$ becomes stronger, and in the strong 
$t$-dimerization limit, $D$ approaches the value 0 at around the 
phase-separation point.  The numerical technique used here is to 
thread the cluster ring with a flux $\Phi$ and study the functional 
form of the ground-state energy with respect to the threaded flux, 
$E_0(\Phi)$.  In general, $E_0(\Phi)$ consists of a series of 
parabola, corresponding to the curves of the individual many-body 
states $E_n(\Phi)$.  This envelope exhibits a periodicity of one 
in units of the flux quantum $\Phi_0$$=$$hc/e$.  
The function $E_0(\Phi)$ (or the Drude weight and superfluid density) 
also yields information on the phenomenon of anomalous flux 
quantization; one may simply include the effect of a constant 
vector potential along the ring by the gauge transformation 
\begin{equation}
\hat{c}_{j\sigma} \to \hat{c}_{j\sigma} e^{ij\phi},~~~
\hat{c}_{j\sigma}^\dagger \to \hat{c}_{j\sigma}^\dagger e^{-ij\phi}
\end{equation}
where
\begin{eqnarray}
\phi=\frac{2\pi}{N_s}\frac{\Phi}{\Phi_0}.  
\end{eqnarray}
The existence of minima at intervals of half a flux quantum, which is 
the anomalous flux quantization, clearly indicates the existence of 
pairing.  In Fig.~7, we show the calculated results for the flux 
quantization.  We see that for $\tilde{t}_d$$=$$0$ (i.e., the uniform 
$t$$-$$J$ model), the anomalous flux quantization is not observed.  
It occurs in the region where the binding energy is negative.  
At any strength of $t$-dimerization we find that the anomalous flux 
quantization occurs for appropriate strength of $J$-dimerization.  

In order to check the consistency of the TL-liquid relations, we 
compare the charge velocity obtained by two independent methods: 
one is from Eq.~(15), and the other is from the relation 
\begin{eqnarray}
v_c = \sqrt{\frac{D}{\pi\kappa}}
\end{eqnarray}
derived from Eqs.~(14) and (17).  The result is shown in Fig.~8, 
where we find that the reasonable consistency is indeed achieved.  

\section{PHASE DIAGRAM}

By summarizing the results for a number of quantities obtained in 
the previous sections, we now draw the phase diagram of the 1D 
dimerized $t$$-$$J$ model at quarter-filling on the parameter space 
of $t$- and $J$-dimerizations.  The result is shown in Fig.~9.	

When the value of $J_2/t_2$ (or the strength of $J$-dimerization) 
is very large, the system is phase separated, which occurs at around 
$J_2/t_2$$\simeq$$3$ when $t_d$$\rightarrow$$0$ and extends to the 
large $\tilde{t}_d$ region.  The critical strength of $J_2/t_2$ is 
almost independent to $\tilde{t}_d$.  The phase separated region is 
determined by the condition $\kappa$$<$$0$.  
When $t$-dimerization is dominant over $J$-dimerization, we find that 
the system becomes a Mott insulator, where the phase boundary is 
determined from the calculated values of the charge gap.  
On the other hand, when $J$-dimerization is dominant over $t$-dimerization, 
the system becomes metallic or superconducting.  We note that there 
always appears the region of the TL-liquid phase between the region 
of the Mott insulating phase and the region of the LE-liquid phase.  
The phase boundary between the TL-liquid and LE-liquid regions is 
determined from the calculated values of the spin gap although whether 
the spin-gap region exists in the homogeneous $t$$-$$J$ model at 
quarter filling is not clear from our exact-diagonalization data\cite{[10]}.  
In the limit of strong $t$-dimerization (i.e., $\tilde{t}_d$$\to$$\infty$), 
the model represents an assembly of isolated dimers and the system 
is an insulator in the entire region of $J_2/t_2$ unless the phase 
separation occurs.  

As for a possible correspondence with experiment, we may refer to a 
Bechgaard salt (TMTTF)$_2$X where the dimerization strength of 
$\tilde{t}_d$$\simeq$$0.1$ is reported and the Mott-insulator to metal 
transition induced by pressure has been observed\cite{[20],[21]}.  
We could argue that our phase diagram includes this phase transition 
at around $\tilde{t}_d$$\simeq$$0.1$ and $J_2/t_2$$\simeq$$0.3$ provided 
that the organic system can be described by the dimerized $t$$-$$J$ 
model with a reasonable range of the parameter values $J/t$.  

We have also examined the phase diagram of the two-dimensional (2D) 
dimerized $t$$-$$J$ model at quarter filling\cite{[22]} and found that 
the phase boundary between the Mott insulating phase and the liquid 
phase has quite similar parameter dependence.  This suggests that, 
irrespective of the spatial dimensions, the same mechanism discussed 
in Sec.~II A operates in the present insulator-metal 
(or insulator-superconductor) transition in the quarter-filled 
dimerized $t$$-$$J$ model.  The main difference between the 1D and 2D 
systems is that in 1D the transition from the Mott insulator is 
to the TL-liquid region whereas in 2D it is to the singlet 
superconducting region.  

\section{CONCLUSION}

We have studied the ground state of the one-dimensional dimerized 
$t$$-$$J$ model at quarter filling by using an exact-diagonalization 
technique on small clusters to calculate the ground-state properties 
such as the charge gap, spin gap, binding energy, Tomonaga-Luttinger-liquid 
parameter, Drude weight, anomalous flux quantization, etc.  
Thereby we have shown that the two types of dimerization, i.e., a dimerization 
of hopping integral (called $t$-dimerization) and a dimerization of exchange 
interaction (called $J$-dimerization), plays an essential and mutually 
competing role in controlling the electronic ground state of the system; 
The $t$-dimerization has the effect leading to the repulsive 
interaction among electrons in a dimer and the $J$-dimerization has the 
effect promoting the spin-singlet formation in a dimer.  
The resulting noticeable features we have obtained are the following: 
(i) The competition between $t$- and $J$-dimerizations induce the 
metal-insulator transition.  (ii) There always appears the region of the 
TL-liquid phase between the region of Mott insulating phase and the 
region of the LE-liquid phase.  (iii) The spin gap and singlet binding 
energy are enhanced by the increase of dimerizations.  
We have summarized the calculated results as the phase diagram in the 
parameter space of dimerizations.  

Finally we would emphasize that the correlated electron systems at 
(and around) quarter filling indeed exhibit interesting properties as we 
have seen and further studies should be pursued in other quarter-filled 
systems from both theoretical and experimental sides.  

\acknowledgments
This work was supported in part by Grant-in-Aid for Scientific 
Research from the Ministry of Education, Science, and Culture of 
Japan.  Financial supports of S.~N. by Sasakawa Scientific Research 
Grant from the Japan Science Society and of Y.~O. by Iketani Science 
and Technology Foundation are gratefully acknowledged.  
Computations were carried out in Computer Centers of the Institute 
for Solid State Physics, University of Tokyo and the Institute for 
Molecular Science, Okazaki National Research Organization.  
 
\begin{figure}
\narrowtext
\caption[]{Schematic representation of the 1D dimerized $t$$-$$J$ model.  
The nearest-neighbor bonds have either $t_1$ and $J_1$ (thin solid line) 
or $t_2$ and $J_2$ (bold line).  The unit cells, each of which contains 
two sites, are indicated by dashed lines.}
\label{fig1} 
\end{figure}
\begin{figure}
\narrowtext
\caption[]{(a) Charge gap $\Delta_{c}/t_2$ as a function of $J_2/t_2$.  
(b) Perturbation estimate of the charge gap compared with 
exact-diagonalization data.}
\label{fig2} 
\end{figure}
\begin{figure}
\narrowtext
\caption[]{Spin gap $\Delta_{s}/t_2$ as a function of $J_2/t_2$.}
\label{fig3} 
\end{figure}
\begin{figure}
\caption[]{Binding energy $\Delta_B/t_2$ as a function of $J_2/t_2$.}
\label{fig4} 
\end{figure}
\begin{figure}
\caption[]{(a) TL-liquid parameter $K_\rho$ as a function of $J_2/t_2$.  
(b) Contour map for $K_\rho$ on the $(t_1/t_2,J_2/t_2)$ plane where 
dotted line is the boundary between the finite spin-gap and gapless 
regions (left panel) and distance $r$ dependence of the charge and 
spin correlation functions where dotted lines are from Eqs.~(11) 
and (12) (right panel).}
\label{fig5} 
\end{figure}
\begin{figure}
\caption[]{Drude weight $D$ as a function of $J_2/t_2$.}
\label{fig6} 
\end{figure}
\begin{figure}
\caption[]{Energy difference $E_0(\phi)$$-$$E_0(0)$ as a function of an 
external flux $\phi$.  The $t$-dimerization $\tilde{t}_d$ increases 
from the top panel to the bottom.}
\label{fig7} 
\end{figure}
\begin{figure}
\caption[]{Charge velocity $v_c$ as a function of $J_2/t_2$.  
The solid lines with symbols represent the value estimated from 
Eq.~(15) and the dashed lines with symbols represent the value 
estimated from Eq.~(20).}
\label{fig8} 
\end{figure}
\begin{figure}
\caption[]{Phase diagram of the 1D dimerized $t$$-$$J$ model at 
quarter filling on the parameter space of $t$- and $J$-dimerizations.  
Dotted line separates the LE-liquid region from the TL-liquid region.  
Contour of constant $\tilde{J}_d$ is shown by dotted lines.}
\label{fig9} 
\end{figure}
\end{multicols}

\begin{references}
\bibitem{[1]}  For a review, see D. J\'erome: {\it Organic Conductors}, 
               ed. J. P. Farges (M. Dekker, New York, 1994). 
\bibitem{[2]}  H. Seo and H. Fukuyama, J. Phys. Soc. Jpn. {\bf 66}, 
               1249 (1997). 
\bibitem{[3]}  K. Penc and F. Mila, Phys. Rev. B {\bf 50}, 11429 (1994). 
\bibitem{[4]}  T. Ohama, H. Yasuoka, M. Isobe, and Y. Ueda, preprint (1998). 
\bibitem{[5]}  P. Horsch and F. Mack, preprint cond-mat/9801316.  
\bibitem{[6]}  M. Imada, Phys. Rev. B {\bf 48}, 550 (1993). 
\bibitem{[7]}  M. Ogata, M. U. Luchini, S. Sorella, and F. F. Assaad, 
               Phys. Rev. Lett. {\bf 66}, 2388 (1991). 
\bibitem{[8]}  H. Yokoyama and M. Ogata, Phys. Rev. Lett. {\bf 67}, 
               3610 (1991). 
\bibitem{[9]}  H. Shiba and M. Ogata, Prog. Theor. Phys. Suppl.  {\bf 108}, 
               265 (1992). 
\bibitem{[10]} M. Nakamura, J. Phys. Soc. Jpn. {\bf 67}, 717 (1998). 
\bibitem{[11]} E. Dagotto: Rev. Mod. Phys. {\bf 66}, 763 (1994). 
\bibitem{[12]} P. Nozi\`eres and S. Schmitt-Rink, J. Low Temp. Phys. 
               {\bf 59}, 195 (1985). 
\bibitem{[13]} H. J. Schulz, Phys. Rev. Lett. {\bf 64}, 2831 (1990); 
               Int. J. Mod. Phys. B {\bf 5}, 57 (1991). 
\bibitem{[14]} N. Kawakami and S. K. Yang, Phys. Lett. A {\bf 148}, 
               359 (1990). 
\bibitem{[15]} H. Frahm and V. E. Korepin, Phys. Rev. B {\bf 42}, 
               10553 (1990). 
\bibitem{[16]} W. Kohn, Phys. Rev. {\bf 133}, A171 (1964).  
\bibitem{[17]} B. S. Shastry and B. Sutherland, Phys. Rev. Lett. {\bf 65}, 
               243 (1990). 
\bibitem{[18]} A. Ferreti, I. O. Kulik, A. Lami, Phys. Rev. B {\bf 45}, 
               5486 (1992). 
\bibitem{[19]} J. Voit, Rep. Prog. Phys. {\bf 58}, 977 (1995).  
\bibitem{[20]} S. S. P. Parkin, C. Coulon, D. J\'erome, 
               J. Phys. C {\bf 16}, L209 (1983). 
\bibitem{[21]} K. Murata, H. Anzai, T. Ukachi, T. Ishiguro, 
               J. Phys. Soc. Jpn. {\bf 53}, 491 (1984).  
\bibitem{[22]} S. Nishimoto and Y. Ohta, preprint (1998). 

\end{references}
\end{document}